\documentclass[twocolumn, secnumarabic, amssymb, nobibnotes, aps, prd]{revtex4-1}

\usepackage{graphicx}
\usepackage{dcolumn}
\usepackage{bm}
\begin{document}

\title{Mitigation of Space-Charge-Driven Resonance and Instability in High-Intensity Linear Accelerators via Beam Spinning}

\author{Yoo-Lim Cheon}
\author{Seok-Ho Moon}
\author{Moses Chung*}
\affiliation{Department of Physics, Ulsan National Institute of Science and Technology, Ulsan 44919, Republic of Korea}
\author{Dong-O Jeon*}
\affiliation{Institute for Basic Science, Daejeon 34047, Republic of Korea}

\begin{abstract}
For modern high-intensity linear accelerators, the well-known envelope instability and recently reported fourth-order particle resonance impose a fundamental operational limit (i.e., zero-current phase advance should be less than $90^{\circ}$).
Motivated by the stability of spinning flying objects, we propose a novel approach of using spinning beams to surpass this limit. We discovered that spinning beams have an intrinsic characteristic that can suppress the impact of the fourth-order resonance on emittance growth and the associated envelope instability.
\end{abstract}

\maketitle


    Spinning flying objects such as American footballs, spinning rockets, and rifled bullets
    are stabilized against small disturbances by maintaining a large angular momentum vector in a specific direction~\cite{Spinning.book}.
    In certain situations, the spin motion counteracts a misaligned thrust to the object~\cite{Spinning.book}.
    Motivated by this long-known stability principle of mechanical systems~\cite{Private},
    we address the following fundamental question, which has not been systematically investigated thus far:
    Can charged-particle beams under strong space-charge effects~\cite{Davidson.book,ReiserBook,Wangler.book,Hofmann.book} in high-intensity linear accelerators (linacs)
    benefit from spinning?
    If beam spinning is in fact an effective countermeasure against such effects,
    it could significantly facilitate the application of intense proton and ion beams to intensity-frontier particle and nuclear physics experiments, fusion material irradiation tests, nuclear waste transmutation, and accelerator-driven subcritical reactors.
    Here, space-charge effects include both coherent instabilities (also called parametric resonances) \cite{Hofmann1982, Chen1994, Lund2004, Li2015, Qiang2018} and incoherent resonances (also called particle resonances) \cite{Jeon1999, Franchetti2003}.

    Recently, Jeon and coworkers \cite{Jeon2009} reported $4\sigma=360^{\circ}$ (or 4:1) fourth-order particle resonance in high-intensity linacs for the first time, and then verified it experimentally~\cite{Groening2009, Jeon2016_2}.
    Here, $\sigma$ is the depressed phase advance per cell.
    Subsequent studies discovered that the fourth-order particle resonance is manifested predominantly over the envelope instability when $\sigma$ is kept constant along the linacs \cite{Jeon2016, Jeon2017, Jeon2018}. In particular, Ref.~\onlinecite{Jeon2018} summarizes the difference between coherent instabilities and particle resonances.
    Further investigations, presented in Ref.~\onlinecite{Lim2020}, established that the general stop band for the $4\sigma=360^{\circ}$ fourth-order particle resonance is $\sigma_{0}>90^{\circ}$ and $\min(\sigma)<90^{\circ}$, where $\sigma_0$ is the zero-current phase advance.
    It is worth noting that the fourth-order resonance stop band is wider than the envelope instability stop band, and the envelope instability is induced following the fourth-order resonance only within the envelope instability stop band in the tune-depression space.

    Certainly,
    it is desirable to overcome these operational limitations associated with space-charge-driven resonances and instabilities.
    It has already been discussed (mainly in the context of high-intensity rings)
    that active suppression of coherent instabilities can be achieved through Landau damping \cite{Shiltsev2017} or nonlinear decoherence \cite{Webb2012}
    using octupoles, feedback dampers, electron lenses, or dedicated nonlinear lattices to provide favorable tune spreads of beam particles.
    These methods are aimed at mitigating coherent instabilities originating from collective perturbations or envelope mismatches.
    However, in high-intensity linacs, even for initially well-matched beams without any external repetitive perturbations, nonlinear space-charge forces can excite higher-order particle resonances.
    In particular, our previous study~\cite{Jeon2016} revealed that for $\sigma_0>90^{\circ}$, the fourth-order resonance is always excited,
    whereas the appearance of envelope instability depends on the lattice and initial matching conditions.
    Accordingly, the fourth-order resonance stop band imposes one of the fundamental operational limits: the zero-current phase advance $\sigma_0$ should be maintained below $90^{\circ}$.

    Hence, in this paper, we investigate whether we can mitigate the $4\sigma=360^{\circ}$ fourth-order particle resonance in linacs
    by introducing the novel concept of beam spinning.
    A spinning beam has a non-zero average canonical angular momentum and exhibits rigid-rotor rotation around the beam propagation axis.
    This scheme is based on two notable achievements in beam physics: i) a rigid-rotor beam equilibrium was obtained for an intense beam propagating through a periodic solenoidal lattice \cite{Chen1997}, and ii) a rotating beam was generated by stripping an ion beam inside a solenoid \cite{Groening2014}.
    To take advantage of the beam spinning effect, we consider an axisymmetric system, in which the canonical angular momentum is conserved.
    We note that many modern low-energy superconducting linacs adopt solenoidal focusing lattices and maintain beam axisymmetry.
    For beam generation, we propose the stripping of $\rm H^-(or~ D^-)$ beams using a thin foil inside a pair of solenoids installed in a medium energy beam transport (MEBT) line,
    and injecting the resultant spinning $\rm H^+(or~ D^+)$ beams into the main linac after proper matching.

    Here, we observe that the stop band and emittance growth of coherent (or envelope) instability following the fourth-order particle resonance
    are indeed reduced for spinning beams.
    This is because the beam mismatch triggered by the fourth-order resonance decreases.
    We present both analytical and multi-particle simulation results to support this argument.
    In our analysis, non-KV Gaussian beams are initially rms-matched to a periodic solenoid focusing channel.

    First, we produce Poincar$\rm\acute{e}$ section plots to observe single-particle trajectories for the $4\sigma=360^{\circ}$ fourth-order resonance
    in the context of the particle-core model \cite{Wangler1998, Qian1995, Gluckstern1994}.
    The evolution of the axisymmetric ($\partial/\partial \theta =0$) transverse beam size $r_b$ with canonical angular momentum is given by the following envelope equation \cite{Chen1997},
    \begin{equation}\label{eq:envelope}
        \frac{d^{2}r_{b}(s)}{ds^{2}} + \kappa_z(s) r_{b}(s) - \frac{K}{r_{b}(s)} - \frac{\epsilon_{T}^{2}}{r^{3}_{b}(s)} = 0,
    \end{equation}
    where $\kappa_z(s)$ is the lattice coefficient as a function of axial coordinate $s$,
    and $\epsilon_{T}$ indicates the rms edge emittance, which is four times the transverse rms emittance, $\epsilon_{\rm{rms}}=\epsilon_{T}/4= \sqrt{ \epsilon_{\rm{th}}^2 + \langle \hat{P_{\theta}} \rangle^2 / 4}$.
    Here, $\epsilon_{\rm{rms}}$ is composed of thermal emittance $\epsilon_{\rm{th}}=\sqrt{\det(\Sigma)}$,
    which is related to the determinant of a $4\times4$ beam matrix $\Sigma$, and
    normalized average canonical angular momentum $\langle \hat{P_{\theta}} \rangle$,
    which represents the statistical average of $\hat{P_{\theta}}=\frac{P_{\theta}}{\gamma\beta mc}$ over the beam distribution.
    The eigen emittances are given by $\epsilon_{1},\epsilon_{2}=\epsilon_{\rm{rms}} \pm  \langle \hat{P_{\theta}} \rangle /2$~\cite{Kim2003, WolskiBook}.
    The space-charge perveance is defined by $K=q\lambda/2\pi\epsilon_{0}\gamma^{3}\beta^{2}mc^{2}$ (in MKS units) with line charge density $\lambda$,
    where $\epsilon_0$ is the vacuum permittivity, and $m$ and $q$ are the particle rest mass and charge, respectively.
    Here, $c$ is the speed of light {\it in vacuum} and $\gamma = (1 - \beta^2)^{-1/2}$ is the relativistic mass factor.
    Even when the beam has non-zero canonical angular momentum, the matched solution for the envelope radius in Eq.~(\ref{eq:envelope}) can be easily determined.


    The equation of motion for radial particle coordinate $r$ with a Gaussian self-field can be simplified as \cite{ReiserBook, Lim2020}
    \begin{equation}\label{eq:EOMr}
        r''(s)  + \kappa_z(s) r(s) - \frac{\hat{P_{\theta}}^2}{r^{3}(s)} - K \frac{1-e^{-\left[ r^{2}(s)/\sigma_{r}^{2}(s)\right]}}{r(s)} =0,
    \end{equation}
    where $\sigma_{r}(s)=r_b(s)/\sqrt{2}$ is the rms beam radius calculated from Eq. (\ref{eq:envelope}) and $' \equiv \frac{d}{ds}$. 
    When $\hat{P_{\theta}}=0$, the non-linear space charge term produces a pseudo octupole that drives 4:1 resonance~\cite{WiedemannBook, LarsProceeding}.
    The third term of Eq.~(\ref{eq:EOMr}) adds an effective repulsive force when the canonical angular momentum is not zero~\cite{ReiserBook}; further, it can alter the space-charge-driven resonance structures (see Appendix).
    We note that $\hat{P_{\theta}}= r^{2} \theta' + (q A_{\theta} r/\gamma \beta mc) $ corresponding to a single particle is determined by the initial conditions,
    where the azimuthal vector potential is $A_\theta = B_z (s) r/2$ to the leading order.

    Under the solenoid magnetic field $B_z$, the equations of motion of a single particle are coupled in the $x$- and $y$-directions in a laboratory frame.
    To eliminate the $x-y$ coupling, one normally performs a transformation from a laboratory to Larmor frame \cite{Davidson.book, ReiserBook}.
    The equation of motion of a single particle in the Larmor frame $(X, Y)$ is then represented as
    \begin{equation}\label{eq:EOMxy}
	    X_{\bot}''(s) + \kappa_z(s) X_{\bot}(s) - K\frac{1-e^{-\left[ r^{2}(s)/\sigma_{r}^{2}(s)\right]}}{r^{2}(s)}X_{\bot}(s) = 0,
	\end{equation}
	where $X_{\bot}=X~\mbox{or}~Y$ and $r^{2}=X^{2}+Y^{2}$.
	The canonical angular momentum is invariant under Larmor transformation and is expressed by $\hat{P_{\theta}} = XY'-YX'$.
	If $\hat{P_{\theta}}$ has a finite value, the nonlinear Gaussian self-fields provide an additional coupling between the $X$ and $Y$ directions
    owing to the $r^{2}$ components in the last term of Eq.~(\ref{eq:EOMxy}), reducing the resonance driving effect (see Appendix).
	For the particle-core model, we assumed that every single particle has the same initial canonical angular momentum,
     such that $\langle \hat{P_{\theta}} \rangle = \hat{P_{\theta}}$, and $\langle \hat{P_{\theta}} \rangle$ as well as the rms emittance remains constant along the lattice periods.

    \begin{figure}
        \centering
        \includegraphics[width=85mm,height=85mm]{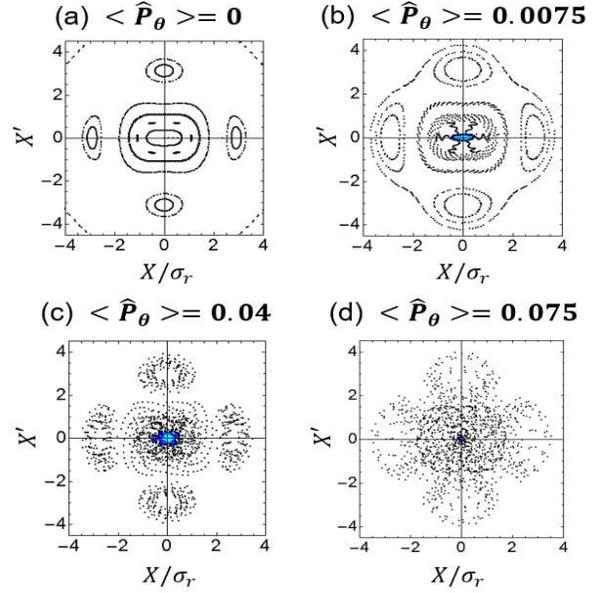}
        \caption{
        Poincar$\rm\acute{e}$ section plots in the Larmor frame when $\sigma_{0}=100^{\circ}$ and $\sigma=72^{\circ}$.
        (a) $\langle \hat{P_{\theta}} \rangle =0$, (b) $\langle \hat{P_{\theta}} \rangle =0.0075$, (c) $\langle \hat{P_{\theta}} \rangle =0.0375$, and (d) $\langle \hat{P_{\theta}} \rangle =0.075$.
        The resonance island structures become blurred as $\langle \hat{P_{\theta}} \rangle$ (in arbitrary units) increases.
        }
        \label{fig:Particle-core_72}
    \end{figure}

    Figure \ref{fig:Particle-core_72} shows the Poincar$\rm\acute{e}$ section plots in the Larmor frame with several different values of the canonical angular momentum when $\sigma_{0}=100^{\circ}$ and $\sigma=72^{\circ}$.
    Here, the same periodic solenoidal lattice $\kappa_z (s) = \kappa_z (s+S)$ as in Ref.~\onlinecite{Lim2020} is employed.
    We note that $\sigma$ is calculated by $\sigma\equiv \epsilon_{\rm rms} \int_{0}^{S}\frac{1}{\langle x^{2} \rangle}ds$,
    where $\sqrt{\langle x^{2} \rangle} = \sqrt{\langle X^{2} \rangle}$ is the rms beam size and $S$ is the lattice period.
    Hence, hereafter, $\sigma$ represents the phase advance of an entire beam distribution, not of a single particle.
    If $\langle \hat{P_{\theta}} \rangle$ equals zero as in Fig.~\ref{fig:Particle-core_72}(a),
    particles have no coupling in the $X$ and $Y$ directions, showing that the four resonance islands are evidently separated from the central region with a tune of 0.25 ($=90^{\circ}/360^{\circ}$).
    By contrast, in the cases of non-zero $\langle \hat{P_{\theta}} \rangle$ beams plotted in Figs.~\ref{fig:Particle-core_72}(b)--(d),
    the coupling effect increases, and the resonance islands become blurred.
    The separatrix and the central elliptical orbits merge,
    which indicates that the resonance particles trapped in the four separate islands enter the stable region and suppress the evolution of halo particles.


    Motivated by the analytical interpretation,
    we perform numerical simulations to obtain clearer evidence for the mitigation phenomena of the fourth-order particle resonance and the associated envelope instability.
    Particularly, we use the TraceWin particle-in-cell code \cite{TraceWin}.

    To generate spinning beams in the simulations, we load initial particles at the center of a solenoid field $B_z$ without any average rotation.
    The total canonical angular momentum is then given by $P_{\theta}=q B_{z} (s=0) r^{2}/2$.
    Outside the magnetic field region, where $B_{z}$ vanishes,
    beam particles gain mechanical angular momentum $P_{\theta}=\gamma\beta mc r^{2} \theta'$.
    This property has been experimentally adopted in research entailing electron beams ~\cite{Burov2000, Sun2004}, and is often referred to as Busch's theorem~\cite{ReiserBook, Burov2000,  Busch2018}.
    We propagate this spinning beam through a periodic solenoid focusing channel with initially well-matched conditions.
    The initial emittance for a non-spinning beam is $\epsilon_{\rm{rms}}=\epsilon_{\rm th} = 6.85$ mm-mrad, and the average initial canonical angular momentum of the spinning beam is calculated over 100,000 particles.
    If we assume $\theta'$ to be the same for all particles, the rotating angle per lattice period would be approximately $30^{\circ}$, $45^{\circ}$, $60^{\circ}$, and $90^{\circ}$
    for $\langle \hat{P_{\theta}} \rangle = 3.75$, 7.5, 11, and 18.7 mm-mrad, respectively.
    The initial rms emittances for the spinning beams are $\epsilon_{\rm{rms}}=7.28$, 7.97, 9, and 11.7 mm-mrad, respectively.
    The longitudinal rms emittance and phase advance are assumed to be extremely small so that the coupling between the transverse and longitudinal dimensions can be ignored.


    \begin{figure}
        \centering
        \includegraphics[width=85mm,height=57mm]{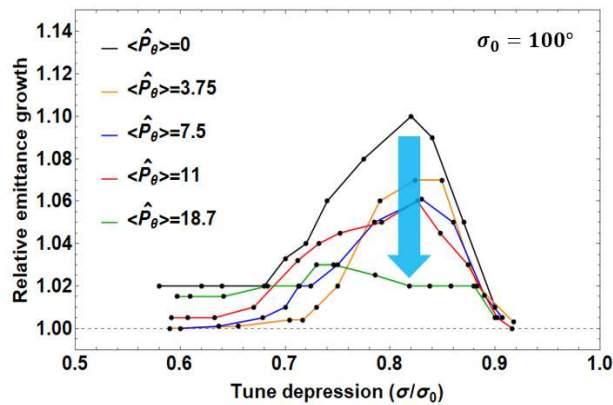}
        \caption{Relative emittance growth during 50 lattice periods for $\sigma_{0}=100^{\circ}$ (i.e., beyond the $90^\circ$ limit), in which the $4\sigma=360^{\circ}$ fourth-order particle resonance is dominant.
        Here, the unit of $\langle \hat{P_{\theta}} \rangle$ is mm-mrad.
        For non-zero $\langle \hat{P_{\theta}} \rangle$ beams, the emittance growth becomes smaller and flat over the tune depression space.}
        \label{fig:EG_1}
    \end{figure}

    \begin{figure}
        \centering
        \includegraphics[width=85mm,height=100mm]{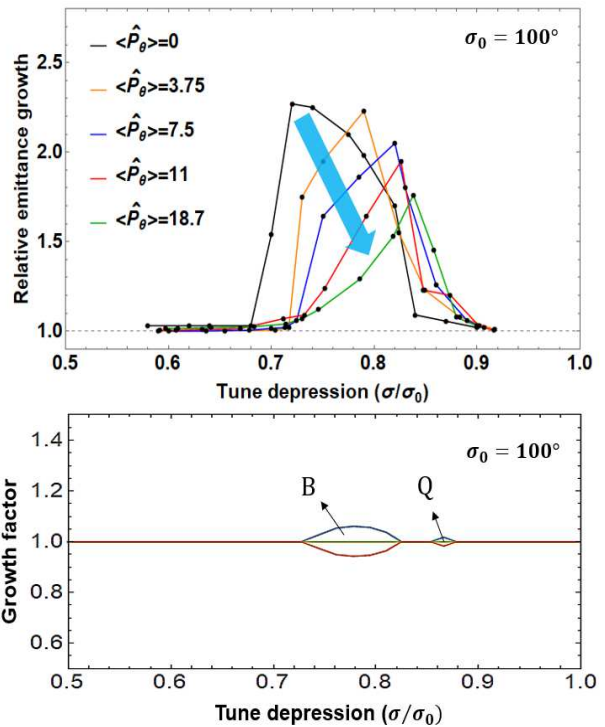}
        \caption{Upper panel shows the plot of the relative emittance growth after 200 lattice periods for $\sigma_{0}=100^{\circ}$
        in which the envelope instability is manifested following the $4\sigma=360^{\circ}$ fourth-order particle resonance.
        Here, the unit of $\langle \hat{P_{\theta}} \rangle$ is mm-mrad.
        Lower panel shows the plot of the envelope instability stop band for $\sigma_{0}=100^{\circ}$, including B-mode and Q-mode instabilities.
        Here, the growth factor corresponds to the amplitude of the eigenvalue of the envelope mode evolution~\cite{Lund2004}.}
        \label{fig:EG_2}
    \end{figure}

    Figure \ref{fig:EG_1} compares the relative emittance growth (the final emittance normalized to the initial emittance) during 50 lattice periods for different values of canonical angular momentum.
    Over such a short propagation, the $4\sigma=360^{\circ}$ fourth-order particle resonance is dominantly manifested against the envelope instability, i.e.,
    the emittance growth is affected by only the fourth-order resonance under well-matched conditions.
    As discussed in Ref.~\onlinecite{Lim2020}, the fourth-order resonance stop band for $\sigma_{0}=100^{\circ}$ is $\sigma/\sigma_{0} \lesssim 0.9$.
    Interestingly, as $\langle \hat{P_{\theta}} \rangle$ increases, the emittance growth becomes smaller and flat over the tune depression space.
    This implies that the emittance growth reaches a certain limit owing to the mitigation of the fourth-order resonance through the nonlinear coupling inherent to the spinning beams (see also Fig.~\ref{fig:Particle-core_72}).

    Figure \ref{fig:EG_2} shows the relative emittance growth over 200 periods for Gaussian beams as well as the envelope instability stop band
    (i.e., the region of the parameter space wherein growth factor $> 1$) of the equivalent KV beam.
    When $\sigma<90^{\circ}$, the fourth-order resonance is excited as depicted in Fig.~\ref{fig:EG_1}, generating a beam mismatch. The beam mismatch excites the envelope instability, leading to significant emittance growth in the envelope instability stop band.
    The stop band of the envelope instability for a fixed $\sigma_{0}$
    is defined as a function of only $\sigma/\sigma_{0}$, independent of $\langle \hat{P_{\theta}} \rangle$~\cite{Hofmann1982, Lund2004}.
    If $\langle \hat{P_{\theta}} \rangle$ = 0, envelope instability is significant, mainly within and near the breathing mode (B-mode) stop band under the periodic solenoid channel.
    The emittance growth is maximum
    near the lower bound of the B-mode stop band and sharply drops to 1 outside the boundary \cite{Lim2020, Hofmann2015}.

    By contrast, the beam mismatch generated by the fourth-order resonance is reduced for the spinning beams,
    which accordingly mitigates the excitation of the envelope instability.
    In the region of $\sigma/\sigma_{0} \leq 0.82$ in Fig.~\ref{fig:EG_1} (left side of the cyan arrow),
    the fourth-order particle resonance is mitigated.
    The beam mismatch and emittance growth are not sufficiently strong to develop the envelope instability, even though the tune depression lies within the instability stop band.
    Therefore, the stop bands of the relative emittance growth in Fig.~\ref{fig:EG_2} shift to the right and become narrower
    as $\langle \hat{P_{\theta}} \rangle$ increases (see the cyan arrow).
    The maximum growth also decreases, and concurrently, the corresponding tune depression value shifts to the right.
    When $\langle \hat{P_{\theta}} \rangle$ is large in the range of $\sigma/\sigma_{0}=0.84\sim0.86$,
    the relative emittance growth in Fig.~\ref{fig:EG_2} is still considerable despite the lower growth compared with the $\langle \hat{P_{\theta}} \rangle =0$ case in the early stage (see Fig.~\ref{fig:EG_1}).
    This is because the beams are affected more by the quadrupole mode (Q-mode) instability owing to the shifted stop bands in the low-beam-current region.
    As the beam current decreases further such that $\sigma/\sigma_{0} > 0.9$, both the fourth-order particle resonance and envelope instability are not observed.

\begin{figure}
    \centering  
    \includegraphics[width=85mm,height=120mm]{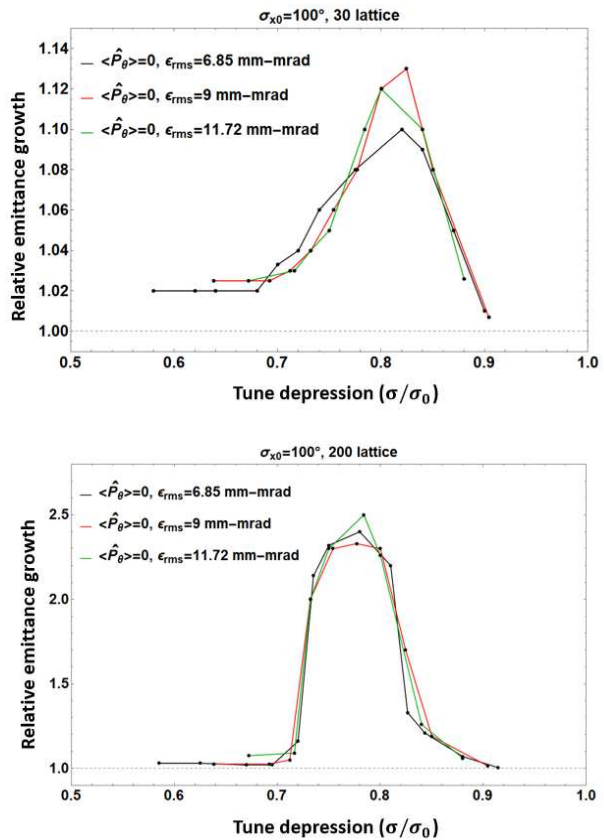}
    \caption{Relative emittance growths after 30 (upper panel) and 200 (lower panel) lattice periods for non-spinning beams with increased initial rms emittance. An increase in rms emittance has a minor influence on the resonance, only reducing the space charge force owing to an increase in beam size.}
    \label{fig:EG_no}
    \end{figure}

    Note that the coupling effect associated with the non-zero canonical angular momentum is evidently different from the case with an increase in thermal emittance ($\epsilon_{\rm th}$) only.
    The particle resonance and envelope instability stop bands are nearly independent of the initial emittances, and there is no mitigation impact in the absence of spinning (see Fig.~\ref{fig:EG_no}).
    Indeed, the initial rms emittance increases by adding the canonical angular momentum and
    by increasing the usual thermal emittance reveal completely different halo population characteristics.

    \begin{figure}
        \centering
        \includegraphics*[width=85mm,height=85mm]{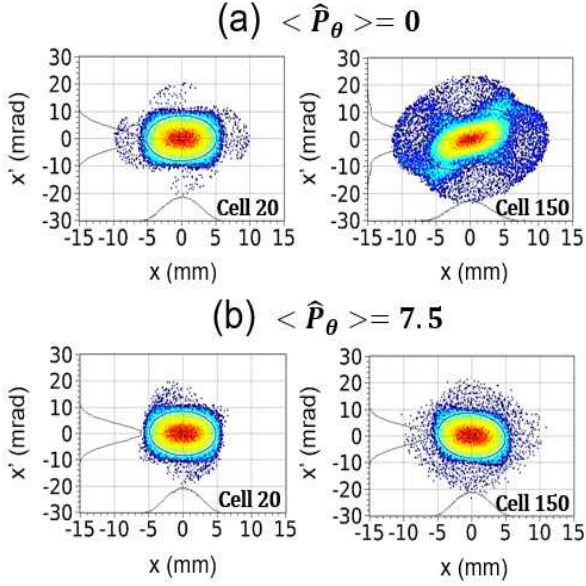}
        \caption{Particle phase space plots in the drift spaces, where the laboratory coordinates are equivalent to the Larmor frame coordinates, for $\sigma_{0}=100^{\circ}$.
         For (a) $\langle \hat{P_{\theta}} \rangle = 0$ and $\sigma=72^{\circ}$,
         the fourfold structure is dominantly observed, which eventually leads to a B-mode instability.
         For (b) $\langle \hat{P_{\theta}} \rangle = 7.5$~mm-mrad and $\sigma=72.5^{\circ}$,
         the fourth-order resonance is mitigated, and envelope instability is not induced even after 150 lattice periods.
         The relative emittance growth is (a) 1.04 and (b) 1.02 at cell 20, and (a) 2.27 and (b) 1.06 at cell 150, respectively.}
        \label{fig:phase_1}
    \end{figure}

    \begin{figure}
        \centering
        \includegraphics[width=85mm,height=85mm]{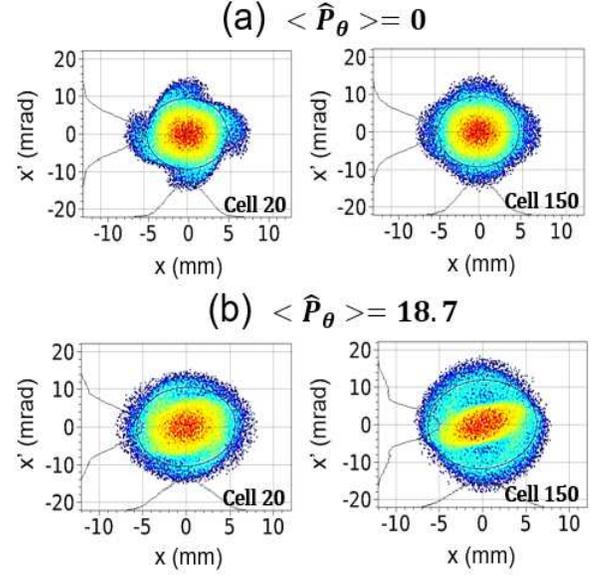}
        \caption{Particle phase space plots in the drift spaces, where the laboratory coordinates are equivalent to the Larmor frame coordinates, for $\sigma_{0}=100^{\circ}$.
         For (a) $\langle \hat{P_{\theta}} \rangle = 0$ and $\sigma=84^{\circ}$,
         the fourth-order resonance persists over 150 lattice periods.
         For (b) $\langle \hat{P_{\theta}} \rangle = 18.7$~mm-mrad and $\sigma=86^{\circ}$,
         the fourth-order resonance is mitigated, but a Q-mode instability occurs after 150 lattice periods.
         The relative emittance growth is (a) 1.09 and (b) 1.02 at cell 20, and (a) 1.09 and (b) 1.45 at cell 150, respectively.}
        \label{fig:phase_2}
    \end{figure}

    The detailed beam distributions in the $x-x'$ phase-space plane are plotted in Fig.~\ref{fig:phase_1}.
    Figure \ref{fig:phase_1}(a) shows the projections of particles at lattice cells 20 and 150 when $\langle \hat{P_{\theta}} \rangle$ equals 0 and $\sigma=72^{\circ}$.
    At cell 20, the resonance particles constitute a fourfold structure propagating outwards from the center, where the relative emittance growth is approximately 1.04, as shown in Fig.~\ref{fig:EG_1}.
    Eventually, the beam attains envelope instability at cell 150 because it is within the stop band.
    The corresponding relative emittance growth becomes approximately 2.27, as can be observed from Fig.~\ref{fig:EG_2}.
    Figure \ref{fig:phase_1}(b) illustrates the projections of the $\langle \hat{P_{\theta}} \rangle = 7.5$~mm-mrad beam when $\sigma=72.5^{\circ}$.
    At cell 20, we clearly observe that the fourth-order particle resonance is mitigated, with fewer particles populating the fourfold structure compared to those in the case of Fig. \ref{fig:phase_1}(a).
    The beam particles gather close to the stable central region with a smaller relative emittance growth of 1.02.
    As a result, the envelope instability is not induced along 150 periods, and the relative emittance growth does not change considerably (from 1.02 to 1.06).

    Figure \ref{fig:phase_2} shows the particle projections for the lower beam currents (i.e., higher $\sigma/\sigma_0$).
    The relative emittance growth in Fig.~\ref{fig:phase_2}(a) is 1.09 at cell 20, and the fourfold structure can be more clearly observed than in the case with the spinning beam.
    Nevertheless, because $\sigma=84^{\circ}$ is outside the stop band of the envelope instability,
    the fourth-order resonance persists over 150 periods, and the relative emittance growth remains at 1.09.
    For the $\langle \hat{P_{\theta}} \rangle = 18.7$~mm-mrad beam in Fig. \ref{fig:phase_2}(b), the relative emittance growth is 1.02 at cell 20 and 1.45 at cell 150.
    As discussed previously, the large-average-angular-momentum beams are affected more by the Q-mode instability for this tune depression range.
    Therefore, the emittance growth at cell 150 becomes larger than that in the case of $\langle \hat{P_{\theta}} \rangle = 0$.
    Nonetheless, it is less adverse than the B-mode instability.


    In summary, we have demonstrated beam spinning as a possible control knob
    for mitigating the fourth-order particle resonance and subsequent envelope instability in modern high-intensity linacs.
    Unlike the other approaches based on nonlinear lattices \cite{Webb2012, Batygin2016},
    which may require a complicated design and reconfiguration of the focusing elements,
    the proposed scheme can be readily applied to high-intensity linacs with periodic solenoidal channels.
    The technology for beam spinning is based on several well-established experiments \cite{Shishlo2012, Groening2014}.
    Using a thin carbon foil (but without a solenoid field) in the MEBT section of the Spallation Neutron Source (SNS),
    a previous study reported the stripping of 2.5 MeV $\rm H^-$ to a proton beam with an efficiency of 99.98\% and emittance growth of only 10--20\% \cite{Shishlo2012}.
    If we strip the $\rm H^-$ beam inside a solenoid field $B_z$ as in the emittance transfer experiment (EMTEX) \cite{Groening2014},
    $\langle \hat{P_{\theta}} \rangle = 2  \kappa_0 \sigma_r^2$ with
    $\kappa_0 =  \left\{ \left[(B\rho)_{\rm in}/(B\rho)_{\rm out} \right]  -  1 \right\} \left[B_z/ 2 (B\rho)_{\rm in} \right]$,
    where $(B\rho)_{\rm in}$ and  $(B\rho)_{\rm out}$ are the beam rigidities before and after the foil, respectively \cite{Chung2016}.
    For a 2.5 MeV $\rm H^-$ beam with $\sigma_r = 2~\rm{mm}$,
    we have $ | \langle \hat{P_{\theta}} \rangle | \lesssim 35$~mm-mrad for $|B_z| \le 1~\rm{T}$.
    For a 2.5 MeV/u $\rm D^-$ beam with similar conditions, we expect an outgoing deuteron beam with $ | \langle \hat{P_{\theta}} \rangle | \lesssim 12$~mm-mrad.
    Hence, the experimentally available ranges of $\langle \hat{P_{\theta}} \rangle$ would cover most of the simulation settings presented in this paper.

\section*{acknowledgements}
    This work was supported by the National Research Foundation (NRF) of Korea (Grant Nos. NRF-2019R1A2C1004862 and NRF-2020R1A2C1010835).
    This work was also supported by the Rare Isotope Science Project of the Institute for Basic Science funded by the Ministry of Science and ICT (MSIT) and
    the NRF of Korea under Contract 2013M7A1A1075764.

*Authors to whom correspondence should be addressed: mchung@unist.ac.kr and jeond@ibs.re.kr

\section*{appendix: A simplified theoretical model for the fourth-order resonance condition}
Here, we qualitatively explain the resonance detuning mechanism both in the cylindrical coordinates and Larmor frame coordinates
using a simplified theoretical model.
First, let us consider the case with $\hat{P}_\theta = 0$.
Using Eq. (2) of the main article, one can expand the nonlinear space-charge term as follows:
\begin{eqnarray}
r'' (s) + \kappa_z (s) r (s)
&=& K \frac{1 - e^{-\left[ r^2 (s) / \sigma_r^2 (s) \right]}}{r (s)} \nonumber \\
&=& K \left[ \frac{r(s) }{\sigma_r^2(s)} - \frac{1}{2}\frac{r^3(s)}{\sigma_r^4(s)} + \cdots  \right].
\end{eqnarray}
Then, we have
\begin{equation}
r''(s) + \left[ \kappa_z (s) - \frac{K}{\sigma_r^2(s)} \right] r(s) = - \frac{K}{2} \frac{r^3(s)}{\sigma_r^4(s)}.
\label{supple2}
\end{equation}
We assume a matched rms beam envelope in an axisymmetric system that is breathing with the lattice period $S$ as~\cite{LarsProceeding}
\begin{equation}
\sigma_r \approx R_0 + \Delta R e^{i 2 \pi s / S}.
\end{equation}
Furthermore, within the smooth-focusing approximation~\cite{Davidson.book}, one may use a constant lattice coefficient
\begin{equation}
\kappa_{sf} \approx \overline{\left[ \kappa_z (s) - \frac{K}{\sigma_r^2(s)} \right]},
\end{equation}
where the overline indicates an appropriate averaging over the lattice period.
Using the definition of the (smooth-focusing) depressed phase advance per cell $\sigma \approx \sqrt{\kappa_{sf}} S$,
one may choose the following form of unperturbed betatron oscillation:
\begin{equation}
r_0(s) \approx C \cos(\sigma s/ S - \phi),
\label{supple5}
\end{equation}
where $C$ is an arbitrary amplitude factor and $\phi$ is an arbitrary phase.
Therefore, the nonlinear perturbation term on the right-hand side of Eq. (\ref{supple2}) scales as
\begin{equation}
- \frac{K}{2} \frac{r^3(s)}{\sigma_r^4(s)} \sim K C^3 e^{\mp 3i \phi} e^{\pm 3 i \sigma s /S} \times \left[ 1 - 4 \frac{\Delta R}{R_0} e^{i 2 \pi s / S} \right].
\end{equation}
Comparing both sides of Eq. (\ref{supple2}), one can obtain the resonance conditions~\cite{WiedemannBook, Chaobook}
\begin{equation}
2 \sigma = 360^\circ  ~~\mbox{or}~~  4 \sigma = 360^\circ.
\label{supple7}
\end{equation}
The first condition $\sigma = 180^\circ$ is simply the stability limit of single-particle motion, which is irrelevant for the present study.
The second condition $\sigma = 90^\circ$ corresponds to the fourth-order resonance condition that is primarily of interest for our discussion.
This fourth-order resonance condition resembles that of octupole perturbation.

When there is a finite canonical angular momentum $\hat{P}_\theta \ne 0$,
the radial coordinate of each particle cannot surpass $r=0$ owing to the additional $\hat{P}_\theta^2/ r^3$ term
in the equation of motion~\cite{ReiserBook}.
Therefore, we cannot use the unperturbed betatron oscillation in the form of Eq. (\ref{supple5}).
In this case, one cannot reproduce the resonance conditions in Eq. (\ref{supple7}).

Since the particle motion in the cylindrical coordinate system is fully equivalent
to that in the Larmor frame, we use the Larmor frame to discuss the detuning from the resonance condition,
without any pitfall related to the $\hat{P}_\theta^2/ r^3$ term in the cylindrical coordinates.
Similar to the analysis in Eq. (\ref{supple2}), one can write the equations of motion in the Larmor frame $(X, Y)$ as
\begin{eqnarray}
X''(s) + \left[ \kappa_z (s) - \frac{K}{\sigma_r^2(s)} \right] X(s) &=& - \frac{K}{2} \frac{r^2(s)}{\sigma_r^4(s)} X(s), \nonumber \\ 
Y''(s) + \left[ \kappa_z (s) - \frac{K}{\sigma_r^2(s)} \right] Y(s) &=& - \frac{K}{2} \frac{r^2(s)}{\sigma_r^4(s)} Y(s).
\label{supple8}
\end{eqnarray}
We note that these equations do not contain the $r^{-3}$ term even when $\hat{P_{\theta}} \ne 0$.
As already discussed in the main article,
we emphasize that if $\hat{P_{\theta}}$ has a finite value, the nonlinear Gaussian self-fields provide an additional coupling between the $X$ and $Y$ directions
through the $r^{2}$ components in Eq.~(\ref{supple8}).
Similar to the ansatz in Eq. (\ref{supple5}),
we choose the following unperturbed solutions in the Larmor frame~\cite{Davidson.book}:
\begin{equation}
X_0(s) \approx C \cos(\sigma s/ S - \phi_1), ~~ Y_0(s) \approx C \cos(\sigma s/ S - \phi_2).
\end{equation}
Here, $\phi_1$ and $\phi_2$ are arbitrary phases.
We note that $\langle X_0^2 \rangle = \langle Y_0^2 \rangle$, as is desired for an axisymmetric beam.
Then, we find the normalized canonical angular momentum to be
\begin{equation}
\hat{P_{\theta}} = X_0 Y_0' - Y_0 X_0' = C^2 \frac{\sigma}{S} \sin(\Delta \phi),
\end{equation}
where $\Delta \phi = \phi_2 - \phi_1$.
For the axisymmetric case ($\partial / \partial \theta =0$) considered in the main article,
$\hat{P_{\theta}}$ remains constant.
Hence, $\Delta \phi$ is also a constant of motion.
For the special case of $\Delta \phi = 0, \pm \pi, \cdots$, we have $\hat{P_{\theta}} = 0$.
For $\Delta \phi = \pm \pi/2, \cdots$, $|\hat{P_{\theta}}|$ becomes maximized.
Finally, the nonlinear perturbation term in Eq. (\ref{supple8}) scales as
\begin{eqnarray}
- \frac{K}{2} \frac{r^2(s)}{\sigma_r^4(s)} X(s) 
&\sim& K C^3 e^{\mp 3i \phi_1} \left[ 1 + e^{\mp 2i \Delta \phi} \right] e^{\pm 3i \sigma s / S} \nonumber \\  
&\times& \left[ 1 - 4 \frac{\Delta R}{R_0} e^{i 2 \pi s / S} \right].
\end{eqnarray}
Therefore, when $\hat{P_{\theta}} =0$ (or, equivalently $\Delta \phi = 0, \pm \pi, \cdots$),
we obtain the same fourth-order resonance condition $4\sigma = 360^\circ$ as before.
By contrast, if $\Delta \phi = \pm \pi/2, \cdots$ (i.e., when $|\hat{P_{\theta}}|$ is maximized),
the detuning term in the above equation $ \left[ 1 + e^{\mp 2i \Delta \phi} \right] =0$, and the fourth-order resonance driving effect (i.e., pseudo-octupole effect) vanishes.
For the ensemble of beam particles with non-zero $\langle \hat{P_{\theta}} \rangle$,
the distribution of $\Delta \phi$ is shifted away from $\Delta \phi = 0, \pm \pi, \cdots$,
and the population of the particles on the fourth-order resonance decreases considerably.
This simplified theory qualitatively explains the observed behavior presented in Fig. 1 of the main article.



\end{document}